\documentclass[aip,jcp,amsmath,amssymb,preprint]{revtex4-1}

\usepackage[dvipdfmx]{graphicx} 
\usepackage{bmpsize}
\usepackage{dcolumn} 
\usepackage{bm} 
\usepackage{here}
\usepackage{color}

\newcommand{\pka}{p$K_A$}
\newcommand{\pkw}{p$K_W$}

\usepackage{amsmath,amsthm,amssymb,cases}

\begin{document}
\title{Nuclear Quantum Effects on Autoionization of Water Isotopologues Studied by Ab Initio Path Integral Molecular Dynamics}

\author{Bo Thomsen}
\email{thomsen.bo@jaea.go.jp}
\affiliation{CCSE, Japan  Atomic Energy Agency, 178-4-4, Wakashiba, Kashiwa, Chiba, 277-0871, Japan}

\author{Motoyuki Shiga}
\email{shiga.motoyuki@jaea.go.jp}
\affiliation{CCSE, Japan  Atomic Energy Agency, 178-4-4, Wakashiba, Kashiwa, Chiba, 277-0871, Japan}
\date{\today}


\begin{abstract}
In this study we investigate the nuclear quantum effects (NQEs)\ on the acidity constant (\pka)
 of liquid water isotopologues at the ambient condition
 by path integral molecular dynamics (PIMD) simulations.
We compared simulations using a fully explicit solvent model with a classical polarizable force field,
 density functional tight binding, and ab initio density functional theory, which
 correspond to empirical, semiempirical, and ab initio PIMD simulations, respectively.
The centroid variable with respect to the proton coordination number of a water molecule
 was restrained to compute the gradient of the free energy, which measures the reversible work
 of the proton abstraction for the quantum mechanical system.
The free energy curve obtained by thermodynamic integration
 was used to compute the \pka\ value based on probabilistic determination.
This technique not only reproduces the \pka\ value of liquid D$_2$O experimentally
 measured (14.86) but also allows for a theoretical prediction of the \pka\ values
 of liquid T$_2$O, aqueous HDO and HTO which are unknown due to its scarcity. 
It is also shown that the NQEs\ on the free energy curve can result
 in a downshift of $4.5\pm 0.9$\ \pka\ units in the case of liquid water,
 which indicates that the NQEs\ plays an indispensable role in the absolute determination of \pka.
The results of this study can help to inform further extensions into the calculation
 of the acidity constants of isotope substituted species with high accuracy.
\end{abstract} 

\maketitle

\section{Introduction}

The acidity constant, \pka, plays a fundamental role in acid-base
 chemistry.
There is no doubt about the importance of quantitatively estimating
 \pka\ for a given functional group in a molecular species.
Computational chemistry rooted in molecular theory has the merit that it is possible
 to evaluate \pka\ independent of experiments.\cite{alongi_chapter_2010,alexov_progress_2011,ho_universal_2010}
However, computational evaluation of \pka\ has yet to reach
 the quantitative level of accuracy, even for the most basic case of liquid water,
 \textit{i.e.}, the water autoionization constant, \pkw.


Previous computational evaluations of \pka\ have been based on either an implicit solvent model,
 in which a solute molecule is embedded in a solvent described
 by polarizable continuum medium,
 \cite{tomasi_quantum_2005,miertus_electrostatic_1981,foresman_solvent_1996,pascualahuir_gepol:_1994,
 miertus_approximate_1982,cossi_energies_2003,klamt_cosmo:_1993,klamt_conductor-like_1995,
 marenich_universal_2009,soteras_extension_2005}
 or an explicit solvent model, in which both the solute
 and the solvent are described explicitly as molecules.
 \cite{doltsinis_theoretical_2003,schilling_determination_2019,chen_prediction_2012,
 davies_estimating_2002,sandmann_copperii-mediated_2019,tummanapelli_dissociation_2014,
 tummanapelli_ab_2015,tummanapelli_ab_2015-1,daub_ab_2019}
In general, the implicit solvent model is able to provide
 \pka\ values with small computational effort but limited
 accuracy.
 The intrinsic error of the implicit solvent model arises from
 the lack of complexity to describe solute-solvent interactions.


 The inclusion of explicit solvent molecules
 \cite{xu_methods_2019,sutton_first_2012,zhang_reliable_2012,thapa_calculations_2015,thapa_density_2016}
  and extended conformational sampling\cite{haworth_modeling_2017}
 have recently been shown to partially solve these issues.
 Both methods,\ however,\ 
 require careful (re)-parametrization of the 
 implicit solvent model and selection of the conformers to be considered in the calculation.
It is therefore preferable to deal with the solvent molecules explicitly
 for quantitatively estimating \pka, where solute-solvent
 interactions are taken into account properly.
In this case \pka\ is estimated directly from
 the free energy change upon the protonation of a solute molecule
 by molecular dynamics (MD) techniques, such as coordination-constrained MD (often to referred to as\
 the blue-moon ensemble method).\cite{carter_constrained_1989,sprik_free_1998}


\pka\ is known to have a strong dependence on
 thermodynamic conditions such as the temperature and pressure.
For liquid water under a pressure of 15 MPa,
 the \pkw\ value of 14 at the ambient temperature decreases
 to about 11 in subcritical conditions at 300 $^\circ$C,
 and then increases to about 20 in supercritical
 conditions at 400 $^\circ$C.\cite{bandura_ionization_2005}
It is also known that \pka\ clearly differs between
 the hydrogen isotopologues,\cite{mora-diez_theoretical_2015}
\textit{e.g.}, \pkw\ of D$_2^{}$O under ambient conditions is 14.86,
 which is larger than that of H$_2^{}$O, 14.00.\cite{shoesmith_ionization_1976}
Constrained MD simulations allow
 the estimation of \pka\ under different thermodynamic conditions.
However, they will produce identical results for
 the hydrogen isotopologues, because the free energy change 
calculated by classical MD  
 does not depend on the nuclear masses.
The isotope effect on the free energy can be traced back to
 the quantum nature whereby the kinetic and potential
 energy operators in the Boltzmann density do not commute.
Therefore, it follows that the nuclear quantum effects (NQEs)\
 should not be ignored for the quantitative estimation of \pka\ .
In fact, hydrogen is generally known to exhibit quantum behaviors such as
 zero-point vibration and tunneling because of its light mass.

The \pkw\ of water has been a long-standing interest for theoretical studies, 
 since the long time-scale on which the autodissociation takes place makes it
  difficult to sample efficiently. 
 Furthermore the strong solvation effects of the produced OH$^{-}_{}$
 and H$_{}^{+}$ ions make it important to consider the solvation structure
 explicitly.

The computation of \pkw\ of liquid water has been done with explicit 
solvent in several studies.\cite{sprik_computation_2000,
perlt_predicting_2017,strajbl_ab_2002,wang_first-principles_2020} 
The reaction mechanism and kinetics 
of the autoionization of liquid water have been the subject in extensive
studies.\cite{moqadam_local_2018,
geissler_autoionization_2001,trout_analysis_1999}
However, none of these studies have explored the NQEs
 on the \pkw\ of water in ab initio simulations based on first principles.


In this paper we introduce a coordination-restrained 
path integral molecular dynamics (PIMD) method
 based on an explicit solvent model to study the \pkw\ of
 liquid water and its isotopologues.
PIMD is a rigorous approach that takes account of
 both nuclear quantum and temperature effects based on
 the imaginary-time path integral theory
 for quantum statistical mechanics 
 \cite{shiga_path_2018,parrinello_study_1984,hall_nonergodicity_1984,
 ceriotti_efficient_2010,tuckerman_efficient_1993}.
PIMD has been used to study the NQEs\ on several properties of water as
 summarized in a recent review.\cite{ceriotti_nuclear_2016}

In this review the NQEs on the \pkw\ of water is estimated by extrapolating the
effect on the water isotopologues to the case of infinite mass nuclei, which
corresponds to the result of a classical MD simulation. The change from
classical to quantum nuclei was found to cause a downshift of around 3 \pka\ 
units. This empirical shift was subsequently used in the study by Wang et al.
to correct the \pkw\ calculated from classical MD.
\cite{wang_first-principles_2020}

 Two important references in the context of this paper are the study
 on the solvated proton and hydroxide ion by Marx \textit{et al.}
 \cite{marx_nature_1999} and the recent study on the NQEs\ in proton transport in water under an electrical field by Cassone.\cite{cassone_nuclear_2020}
Here we will extend the PIMD method 
 for \pkw/\pka\ estimations,
 allowing the computations of quantum free energies upon the
 protonation of the solute, by restraining the centroid variable
 of the coordination number (CN).


In Section II we will outline the theory of 
 coordination-restrained PIMD and
 the calculation of \pkw\ in terms of probabilistic and absolute methods.
Section III contains the computational details for the simulations
 conducted in this study.
The results of MD and PIMD will then be discussed in Section IV.
Furthermore we will investigate the isotope effects
 in pure D$_2$O and T$_2$O, and the solvated HDO and HTO
 isotope substituted water like molecules.
Finally in Section V we will summarize our findings.


\section{Theory}
\subsection{Coordination Number Restrained Path Integral Molecular Dynamics}
Consider the quantum Hamiltonian for an $N$ atom system within the Born Oppenheimer approximation
\begin{equation}
\hat{H}=\sum_{I=1}^{N}\frac{\hat{\mathbf{P}}^2_I}{2M_I}+V(\hat{\mathbf{R}}_1,\ldots,\hat{\mathbf{R}}_N),
\end{equation}
where $M_I$ is the mass of the $I$th particle, $\hat{\mathbf{P}}^2_I$ is the momentum operator 
$\left(\hat{P}_{I,x},\hat{P}_{I,y},\hat{P}_{I,z}\right)$, $\hat{\mathbf{R}}_I$ is the position operator 
$\left(\hat{R}_{I,x},\hat{R}_{I,y},\hat{R}_{I,z}\right)$ and $V$ is the potential. 
The quantum partition function for this system is given as
\begin{equation}
Z=\int d\mathbf{R}_1\cdots \int d\mathbf{R}_N
\left\langle \mathbf{R}_1\cdots \mathbf{R}_N\left| \exp\left(-\beta\hat{H}\right)\right| 
\mathbf{R}_1\cdots \mathbf{R}_N\right\rangle=\mathrm{Tr} \exp\left(-\beta \hat{H}\right).\label{eq::ExactPart}
\end{equation}
Here $\beta=\frac{1}{k_bT}$, where $k_b$ is the Boltzmann constant and $T$ is the temperature. 
It is assumed that the thermal de Broglie wavelength is smaller than the separation of two identical atoms,
 thereby allowing us to ignore the possibility for exchange of fermion/boson positions. 


By dividing the Boltzmann\ operator in Equation (\ref{eq::ExactPart}) into $P$ terms, 
inserting closure relations in coordinate spaces between each term of the Boltzmann\ operator, 
and applying the second-order Suzuki-Trotter expansion one can derive
 the following expression for the partition function,
\begin{equation}
Z=\lim_{P\rightarrow \infty}Z_P
\end{equation}
where
\begin{eqnarray}
Z_P&=&\prod_{I=1}^N\left[ \left(\frac{M_IP}{2\pi \beta\hbar^2}\right)^{\frac{3P}{2}}
\int d\mathbf{R}_I^{(1)}\int d\mathbf{R}_I^{(2)}\cdots \int d\mathbf{R}_I^{(P)}\right]\nonumber\\
&&\times \exp\left[-\beta\left\{\sum_{s=1}^P\sum_{I=1}^N
\frac{M_I}{2}\omega_P^2\left(\mathbf{R}_I^{(s)}-\mathbf{R}_I^{(s-1)}\right)^2+
\sum_{s=1}^P\frac{1}{P}V\left(\mathbf{R}_1^{(s)}, \ldots, \mathbf{R}_N^{(s)}\right)\right\}\right].
\end{eqnarray}
The constant $\omega_P$ is given as $\frac{\sqrt{P}}{\beta\hbar}$. 
The factor $\prod_{I=1}^N\left(\frac{M_IP}{2\pi \beta\hbar^2}\right)^{\frac{3P}{2}}$ is a constant
 and will be omitted in the following, as it does not alter the relative free energy differences. 
Rearranging the exponent in the above equation we arrive at the following,
\begin{equation}
Z_P\propto \prod_{I=1}^N\left[ \int d\mathbf{R}_I^{(1)}\int d\mathbf{R}_I^{(2)}\cdots \int d\mathbf{R}_I^{(P)}\right]
\exp\left(-\beta V_{\mathrm{eff}}(\left\{\mathbf{R}\right\})\right),\label{eq::proppart}
\end{equation}
where
\begin{equation}
V_{\mathrm{eff}}\left(\left\{\mathbf{R}\right\}\right)=
\sum^P_{s=1}\left\{\sum^N_{I=1}\frac{M_I}{2}\omega_P^2
\left(\mathbf{R}_I^{(s)}-\mathbf{R}_I^{(s-1)}\right)^2+\frac{1}{P}V\left(\mathbf{R}_1^{(s)}, 
\ldots, \mathbf{R}_N^{(s)}\right)\right\}.
\end{equation}
These two equations show that the quantum behaviour of an $N$ particle system can be mimicked by 
considering an $NP$ particle system. 
The individual terms in $P$ are often referred to as beads on a chain, where each bead corresponds to a
 single copy of the classical system. 
This system is evolved according to its classical forces and its coupling to a polymer chain. 
The cost of this method thus scales steeply with the number of beads, 
 and by extension accuracy desired. 
Convergence with respect to number of beads does however scale inversely with temperature, 
 \textit{i.e.} a few tens of beads can be sufficient to accurately describe quantum effects 
 for systems under standard conditions.


The blue-moon sampling approach\cite{carter_constrained_1989,sprik_free_1998} 
 is here used to compute the free energy curve of water autoionization. 
Following a study by Sprik\cite{sprik_computation_2000},
 the CN of an oxygen atom (labeled as ``O$^*$'') is used for studying this reaction.
We introduce a rational function for the CN as
\begin{equation}
n_{\mathrm{O^*}}(\{{\bf R}\})=
\sum_{j\in H}\frac{1-\left(\frac{r_{\mathrm{O^*}j}}{d_{\mathrm{OH}}}\right)^6}{1-\left(\frac{r_{\mathrm{O^*}j}}{d_{\mathrm{OH}}}\right)^{12}},
\label{eq::restrain}
\end{equation}
where $\{{\bf R}\}$ is the set of atomic coordinates of the system, 
$r_{ij}$ are the distances to all hydrogen atoms in the system, and $d_{\mathrm{OH}}$ is a constant 
set to 1.35 \AA.
%


The CN is in this study is restrained to vary the coordination of hydrogen in a single OH$^{-}$ moiety
from one to zero. 
These conditions correspond to the moiety forming a H$_2$O molecule and 
a solvated OH$^{-}$ ion respectively.
In practice, for a random oxygen atom, one of its attached hydrogens is chosen to remain bound, 
while the other attached hydrogen and all other hydrogens are subject to the CN restraint. 
The same method is used in the case of D$_2$O and T$_2$O for the deuterium and tritium atoms in the simulation,
 respectively.
For HDO and HTO in solution all hydrogens are restrained, while the core OD$^-$ or OT$^-$ is kept unrestrained.


To derive the free energy difference in the blue-moon ensemble 
one has to consider the free energy of a system restrained 
to a fixed coordination number $\bar{n}_{\mathrm{O^*}}$,
\begin{equation}
A(\bar{n}_{\mathrm{O^*}})=-\beta^{-1}\log \rho(\bar{n}_{\mathrm{O^*}})+A_0.
\end{equation}
$A_0$ is a constant term taking care of the constants dropped in Equation (\ref{eq::proppart}). 
This term is only necessary to consider when calculating the absolute value of the free energy, 
as it will cancel for relative free energy differences. 
The distribution
$\rho(\bar{n}_{\mathrm{O^*}})$ is the scaled probability for finding the system with the given coordination number 
$\bar{n}_{O^*}$, which is expressed as
\begin{eqnarray}
&&\rho(\bar{n}_{\mathrm{O^*}})=\left\langle \delta\left(\bar{n}_{\mathrm{O^*}}-\frac{1}{P}\sum_{s=1}^P n_{\mathrm{O^*}}
\left(\mathbf{R}^{(s)}\right)\right)\right\rangle\nonumber\\
&=&\lim_{P\rightarrow\infty}Z_P^{-1}\prod_{I=1}^N\left[\int d\mathbf{R}_I^{(1)}\cdots \int d\mathbf{R}_I^{(P)}\right]\delta\left(\bar{n}_{\mathrm{O^*}}-\frac{1}{P}\sum_{s=1}^P n_{\mathrm{O^*}}
\left(\mathbf{R}^{(s)}\right)\right)\exp\left(-\beta V_{\mathrm{eff}}(\left\{\mathbf{R}\right\})\right)\nonumber\\
\end{eqnarray}
in the PIMD formalism.
The brackets indicate the ensemble average of 
the contained function,  which is equal to its time average 
assuming ergodicity.
To further simplify the expression a narrow peaked Gaussian function, 
with an inverse variance given as $\frac{\kappa}{2}$, is used to approximate the delta function, giving
\begin{equation}
\rho(\bar{n}_{\mathrm{O^*}}) \approx \lim_{P\rightarrow\infty} Z_P^{-1} \sqrt{\frac{\beta\kappa}{2\pi}} \prod_{I=1}^N
\left[\int d\mathbf{R}_I^{(1)}\cdots \int d\mathbf{R}_I^{(P)}\right] 
\exp\left(-\beta V_{\mathrm{eff}}^{\mathrm{cons}}(\left\{\mathbf{R}\right\},\bar{n}_{\mathrm{O^*}})\right)
\label{eq::ConsFreeEn}
\end{equation}
where
\begin{equation}
V^{\mathrm{cons}}_{\mathrm{eff}}(\left\{\mathbf{R}\right\},\bar{n}_{\mathrm{O^*}})
=V_{\mathrm{eff}}(\left\{\mathbf{R}\right\})+
\frac{\kappa}{2}\left(\bar{n}_{\mathrm{O^*}}-\frac{1}{P}\sum_{s=1}^Pn_{\mathrm{O^*}}
\left(\mathbf{R}^{(s)}\right)\right)^2
\label{eq::ConsFreeEn2}.
\end{equation}
It is exceedingly difficult to calculate the free energy value itself through the above equations. 
Calculating the derivative of the free energy with respect to the restrained value is, however, 
less of a challenge. 
This derivative is given as
\begin{equation}
f(\bar{n}_{\mathrm{O^*}})=\frac{\partial A(\bar{n}_{\mathrm{O^*}})}{\partial \bar{n}_{\mathrm{O^*}}}=
\left\langle \kappa\left(\bar{n}_{\mathrm{O^*}}
-\frac{1}{P}\sum_{s=1}^Pn_{\mathrm{O^*}}\left(\mathbf{R}^{(s)}\right)\right)\right\rangle_{\mathrm{eff}}
\label{eq::resforceav},
\end{equation}
where the subscript ``eff'' stands for the sampling by a PIMD simulation with the restraint.
This derivative corresponds to the time-averaged force which will be used in the following 
to calculate the free energy surface as a function of the CN.

\subsection{Methodology for Calculating the Autoionization Constant of Water}

Using the time-averaged forces, $f(\bar{n}_{\mathrm{O^*}i})$, from Equation (\ref{eq::resforceav}) 
and the time-averaged CNs, $n_{\mathrm{O^*}i}$, obtained from a number 
of CN restrained simulations it is possible to estimate the 
free energy difference through thermodynamic integration,
\begin{equation}
\Delta A(\bar{n}_{\mathrm{O^*}i}) = \int_{\bar{n}_{\mathrm{O^*}1}}^{\bar{n}_{\mathrm{O^*}i}} 
f(\bar{n})d\bar{n}
\label{eq::FreeEnergy}.
\end{equation}
The numerical integral is evaluated by spline interpolation between each of the simulated 
restrained CNs $(\bar{n}_{\mathrm{O^*}1},\ldots,\bar{n}_{\mathrm{O^*}M})$.
The numerical integral is here calculated using a third order B-spline which passes through 
all the calculated points.


Using the free energy surface one can employ a probabilistic (PROB) method to calculate the \pkw\ 
of water and \pka\ of an acid, as suggested by Davies \textit{et al.},\cite{davies_estimating_2002}
based on the work of Chandler.\cite{chandler_introduction_1987}
This method relies on the relative probability of finding the system in a bound state. 
For this purpose we define a cutoff bond distance, $R_c$,
 at which the O-H bond breaks and the OH$^-$ and H$_3$O$^+$ ions are formed.
The probability ratio between the bound and dissociated states is given by
\begin{equation}
\gamma(R_c)=\frac{\int_0^{R_c}\exp(-\beta\Delta A(\bar{n}_{\mathrm{O^*}}(r)))4\pi r^2 dr}
{\int_0^{R_{\mathrm{max}}}\exp(-\beta\Delta A(\bar{n}_{\mathrm{O^*}}(r)))4\pi r^2 dr}\label{eq::prop},
\end{equation}
where the factor $4\pi r^2$ arises from the Jacobian of polar coodinates.
The mapping $\bar{n}_{\mathrm{O^*}}(r)$ is carried out by assigning the time averaged 
closest distance between the central oxygen and the restrained hydrogens, $r_i$, 
for the given restraint $\bar{n}_{\mathrm{O^*}i}$. 
The assigned free energy difference calculated in Equation (\ref{eq::FreeEnergy}), 
$\Delta A(\bar{n}_{\mathrm{O^*}}(r))$, can then be linearly interpolated to numerically evaluate 
the probability. 
$R_{\mathrm{max}}$ is the time averaged distance when the CN
 restraint is set to the lowest coordination number, $\bar{n}_{\mathrm{O^*}M}$.


The value of $R_c$ is determined so
 that it gives $\mathrm{p}K_W=14.00$ for liquid H$_2^{}$O\@.
For water autoionization
\begin{equation}
\mathrm{H_2O(l)}\rightleftharpoons \mathrm{OH^-(aq)+H^+(aq)},\label{eq::waterdiss}
\end{equation}
the autoionization constant is
\begin{equation}
\mathrm{p}K_W=-\log\left([\mathrm{OH^-(aq)}][\mathrm{H^+(aq)}]\right).\label{eq::waterautoion}
\end{equation}
Rewriting this using the probabilities of water dissociation $(\gamma_W(R_c))$ from Equation (\ref{eq::prop}) leads to
\begin{equation}
\mathrm{p}K_W(R_c)=-2\log\left(\left(1-\gamma_W(R_c)\right)\frac{N_W}{c_0V}\right)\label{eq::pkwprop},
\end{equation}
where $N_W$ and $V$ are the number of water molecules and the volume of the simulation box, respectively,
and $c_0$ is the standard concentration (1 M). 
To calculate $R_c$, the reference value at standard conditions is used, resulting in
\begin{equation}
\gamma_W\left(R_c\right)=1-\frac{10^{-7}}{c_W}
\end{equation}
where $c_W=\frac{N_W}{c_0V}$ (which is about 55.6 in ambient conditions).
This can then be used to find $R_c$ for Equation (\ref{eq::prop}).
It is then commonly assumed that this distance is also applicable to breaking the A-H bond
 to form A$^-$ and H$_3$O$^+$ for a common acid A.


The \pka\ of an acid, or acid group of a molecule, (A) in water can be found by considering
\begin{equation}
\mathrm{AH(aq)}\rightleftharpoons \mathrm{A^-(aq)+H^+(aq)},
\end{equation}
and the following expression for the acidity constant (\pka ),
\begin{equation}
\mathrm{p}K_A=-\log\left(\frac{[\mathrm{A^{-}}][\mathrm{H^+}]}{[\mathrm{AH}]}\right).
\label{eq::acidpka}
\end{equation}
Here we have assumed a dilute aqueous solution, where the activities of all species are 
determined by their concentration in the solution.
The solvated protons, $\mathrm{H^+}$, can in principle stem from both water autoionization and 
the dissociation from A leaving us with the following expression for their concentration
\begin{equation}
[\mathrm{H^+}]=[\mathrm{H^+}]_{\mathrm{acid}}+[\mathrm{H^+}]_{\mathrm{water}}
=(1-\gamma_A(R_c))c_A+(1-\gamma_W(R_c))c_W^\prime,
\end{equation}
where $c_W^\prime=\frac{N_W^\prime}{c_0V}$ and $c_A=\frac{N_A}{c_0V}$. $N_W^\prime=N_W-N_A$, and
$N_A$ is the number of acid molecules replacing water molecules in the box.
It is implicitly assumed that the \pkw\ of water is unchanged in the solution of AH,
a reasonable assumption given that the solution is dilute.
Using the probabilities of dissociation for the acid and water we can rewrite Equation (\ref{eq::acidpka}) as
\begin{eqnarray}
\mathrm{p}K_A({\rm PROB})&=&-\log\left(\frac{(1-\gamma_A(R_c))c_A
\left((1-\gamma_A(R_c))c_A+(1-\gamma_W(R_c))c_W^\prime\right)}{c_A\gamma_A}\right)\nonumber\\
&=&-\log\left(\frac{(1-\gamma_{A})}{\gamma_{A}}
\left(\frac{(1-\gamma_{A})N_A}{c_0V}+10^{-7}\frac{N_W-N_A}{N_W}\right)\right).
\label{eq::pkapropext}
\end{eqnarray}
Generally one would find that $(1-\gamma_A(R_c))\gg(1-\gamma_W(R_c))$, so 
the above equation can then be reasonably approximated as
\begin{equation}
\mathrm{p}K_A({\rm PROB})\approx -\log\left(\frac{(1-\gamma(R_c))^2}{(\gamma(R_c))}\frac{N_A}{c_0V}\right)\label{eq::pkaprop}.
\end{equation}

Equation (\ref{eq::pkaprop}) was used in previous studies
 to predict the \pka\ of acidic substances
 where the proton concentration are mainly from the solute dissociation.
\cite{doltsinis_theoretical_2003,schilling_determination_2019,
chen_prediction_2012,davies_estimating_2002,sandmann_copperii-mediated_2019}
In this study we will resort to using Equation (22), since the \pka\ of 
HDO and HTO in aqueous solution is expected to be close to that 
of the solvent H$_2$O itself. 
This approach is however general to all very weak acids, with 
\pka\ values around their solvent. 
These types of acids have not previously been studied in this context, 
which is why most studies have relied on Equation (\ref{eq::pkaprop})
for calculating \pka\ values.


Another way of calculating the probabilities used in the equations above is to employ a basic two state model. 
That is, we assume that the minimum of the free energy surface
 corresponds to the free energy of the bound state of water and the 
 free energy of the lowest CN restraint corresponds to the dissociated state. 
From this model we can formulate the probability of finding a water molecule
in a dissociated state as
\begin{equation}
1-\gamma = \frac{\exp\left(-\beta\Delta A\right)}{1+\exp\left(-\beta\Delta A\right)},
\end{equation}
where $\Delta A$ is taken as the difference between
 the maximum (dissociated state) value
 and the minimum (bound state) value of
 $\Delta A(\bar{n}_{\mathrm{O^*}i})$ in Equation (\ref{eq::FreeEnergy}).
For the autoionization of water and its isotopologues\ we can approximate 
this probability as $\exp\left(-\beta\Delta A\right)$, as the free energy
 difference is very large.
Inserting this into Equation (\ref{eq::pkwprop}) results in 
\begin{equation}
\mathrm{p}K_W({\rm ABS})=-2\log\left(\exp\left(-\beta\Delta A\right)\frac{N_W}{c_0V}\right)=\frac{2\beta\Delta A}{\ln(10)}-2\log\left(\frac{N_W}{c_0V}\right)\label{eq::pwkabs}.
\end{equation}
This method allows us to compare the calculated \pkw\ of H$_2^{}$O
 to that of D$_2$O and T$_2$O, as opposed to the probabilistic method, 
 in which the H$_2^{}$O simulation is used as a reference.
In the following this method will be referred to as the absolute (ABS) method, as
it depends on the absolute free energy difference between two states.
\section{Computational Details}
\begin{figure*}
\includegraphics[width=\linewidth]{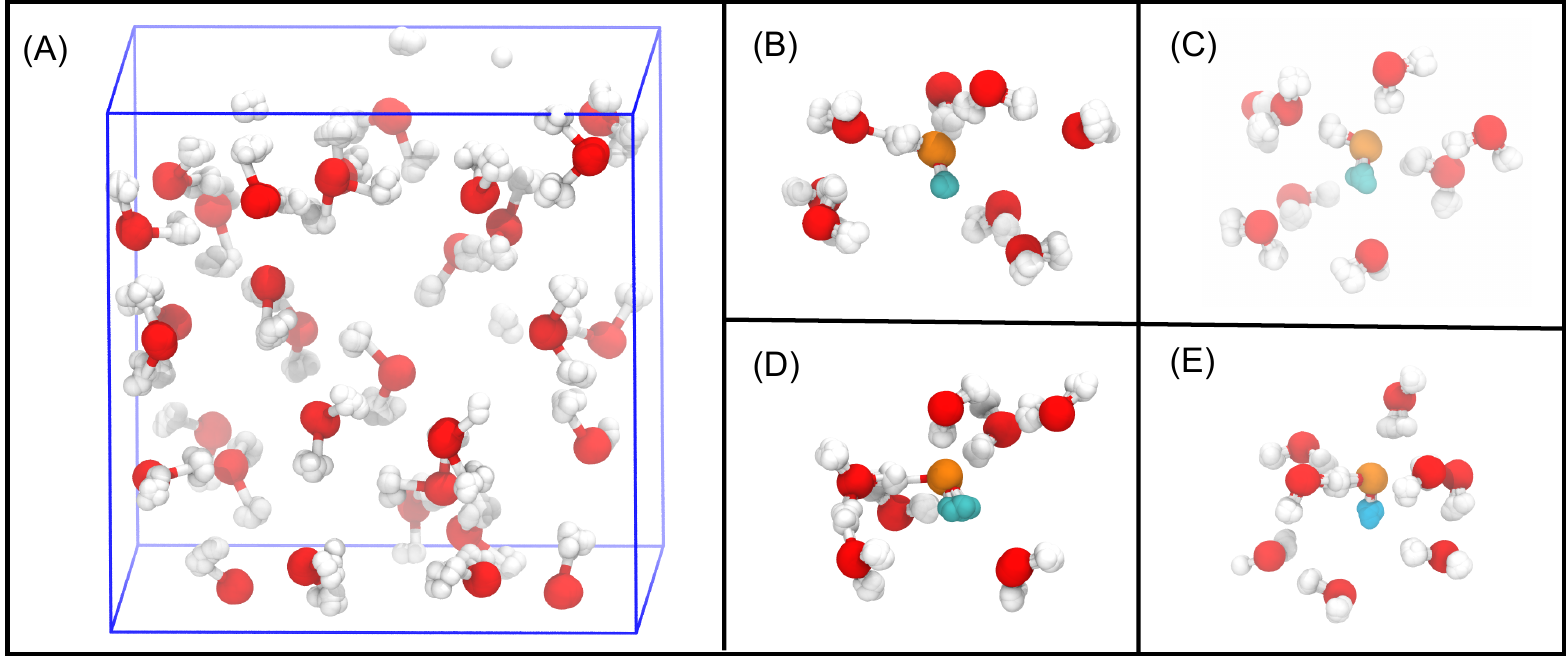}
\caption{}
\label{Figure::SystemSetup}
\end{figure*}

MD and PIMD simulations of liquid H$_2^{}$O and its isotopologues were 
undertaken in the canonical ensemble at temperature 300 K. 
 The interatomic potential and the associated force were computed 
 on the fly by ab initio DFT, semiempirical DFTB, or empirical OSS2 methods. 
 The simulation conditions are listed in Table I. 
Systems of $32-64$ water molecules were contained in cubic boxes with 
 periodic boundary conditions.
 The box size was set such that the number density was 29.86 molecules/\AA$^3$, 
 which amounts to 1.00 g/cm$^3$ for H$_2^{}$O.
An example structure is depicted in Figure \ref{Figure::SystemSetup} 
along with snapshots of the constrained water molecule and its 
surroundings under different CN restraints recorded from the DFT trajectories.


The numerical integration schemes for MD and PIMD were based on the reversible 
reference system propagation algorithm (RESPA) as implemented in the
\texttt{PIMD} software package.\cite{shiga_pimd_2020}
The temperature was strongly controlled by attaching massive Nos\'e-Hoover 
chain (MNHC) thermostats\cite{nose_unified_1984,hoover_canonical_1985,martyna_nosehoover_1992} 
 to each degree of freedom in the MD simulations. 
The MNHC thermostats were also attached to each normal mode representing 
the ring polymer in the PIMD simulations. 
The fifth-order Suzuki-Yoshida factorization was used for the numerical 
integration of the MNHC thermostats. 
The simulations were run\ for $12.5-25.0$ ps with the step size of $0.25-0.43$ fs 
depending on the system and the interatomic potential. 
The restraints with a force constant $\kappa = 4-10$ hartree were applied at 
15 points in the range $0.16 \le n_i^{} \le 0.98$.
This resulted in a fluctuation of CN within the order of 0.01. 
To deal with the fast oscillation caused by the restraints, the restraint force 
was updated 5 times per MD/PIMD step in the RESPA technique. 


The ab initio DFT energy calculations were carried out using the 
Vienna ab initio simulation package (\texttt{VASP}).
\cite{kresse_ab_1993,kresse_ab_1994,kresse_efficiency_1996}
We employed the Perdew-Burke-Ernzerhof (PBE)\cite{perdew_generalized_1996}
exchange correlation functional and Grimme’s D3 van der Waals
correction.\cite{grimme_consistent_2010}
The core electrons were taken into account using the projector augmented wave
(PAW) method.
The valence electrons are expressed in terms of a linear combination of plane
wave basis functions with a cutoff at 400 eV.
Only the $\Gamma$-point of the Brillouin zone was computed.


The semiempirical DFTB energy calculations were carried out using 
\texttt{DFTB+}.\cite{hourahine_dftb+_2020}
We employed the third-order self-consistent-charge density-functional 
tight-binding (SCC-DFTB)\cite{elstner_self-consistent-charge_1998} 
method with the 3ob Slater-Koster parameter set.\cite{gaus_parametrization_2013}


The empirical energy calculations based on the OSS2\cite{ojamae_potential_1998} 
potential were implemented in an in-house version of the \texttt{PIMD} software package. 
The OSS2 potential can be categorized as a polarizable force field composed of 
short-range intramolecular bonds and long-range interactions between point charges 
and induced point dipoles with damping. 
The parameters in the OSS2 potential are fitted so as to reproduce the ab initio 
energies of neutral and protonated water clusters at the level of second-order 
M\o ller-Plesset perturbation theory (MP2). 
The OSS2 potential was originally developed for small water clusters in 
the free boundary condition, but it can be applied to bulk water by 
adaption for periodic boundary conditions. 
We used the Ewald sum of point charges and induced point 
dipoles for this, where the point dipoles in the electrostatic field were determined 
by the matrix inversion method.\cite{sala_polarizable_2010}


The resulting restraints, forces and trajectories were analysed using a locally 
developed python script, where the \texttt{MDAnalysis} library
\cite{michaud-agrawal_mdanalysis:_2011,gowers_mdanalysis:_2016} 
was used for calculating the O-H distances. 
The errors reported here were calculated by the method outlined by Flyvbjerg 
and Petersen.\cite{flyvbjerg_error_1989}
The errors of the cut off distance $R_c$ and the probabilistic \pka\ or \pkw\ are 
not considered to be correlated, \textit{i.e.}, the errors in \pkw\ or \pka\ are 
calculated using a fixed value of $R_c$. 
All figures depicting the molecular systems were visualized using the \texttt{VMD}
software.\cite{humphrey_vmd:_1996} 
The numerical integrals needed for the calculations outlined in the theory section
were all carried out using linear interpolation and a step size of $1\cdot 10^{-4}$
in both CN and O-H distance space.

\section{Results}

\begin{figure}
\includegraphics[width=0.9\linewidth]{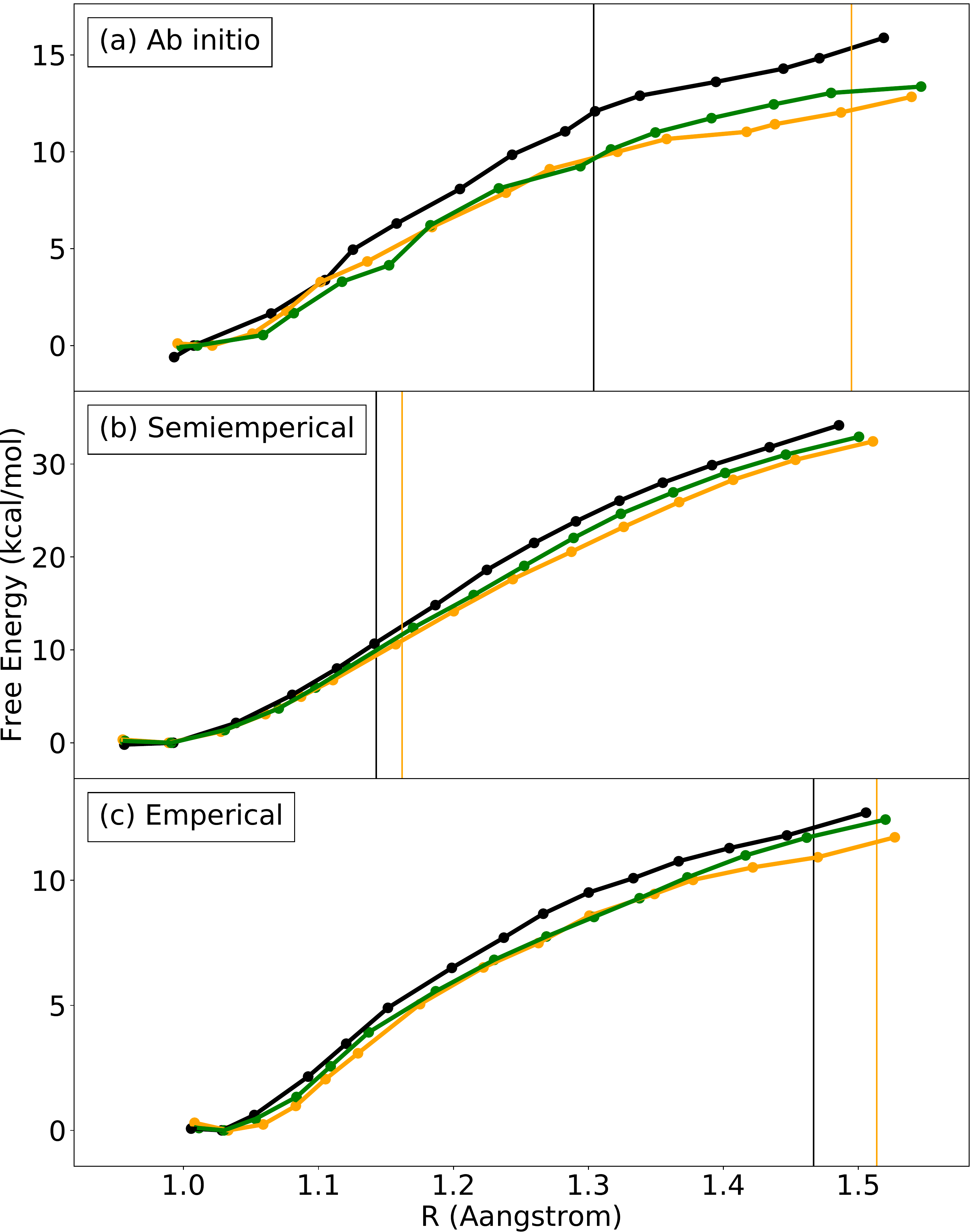}
\caption{}
\label{fig::FreeEnerg}
\end{figure}

The free energy curves, $A(n(r))$, obtained from ab initio MD and PIMD simulations 
are displayed in Figure \ref{fig::FreeEnerg}(a).
These represent in our opinion the most reliable results in the present study.
The following features become clear when comparing the results presented in 
Figure \ref{fig::FreeEnerg}(a).
The ab initio PIMD results in a much lower free energy for the dissociation 
process when compared to that of ab initio MD.
This effect can be explained by the NQEs, which is expected to lead to lower
free energies of dissociation due to the delocalization of the proton.
\cite{ceriotti_nuclear_2016}
Comparing the results for the isotopologues H$_2^{}$O and T$_2^{}$O we find that
the free energy curves are different in the ab initio PIMD. 
The calculation of these two species using ab initio MD would result in
the same curves due to the absence of NQEs in classical MD
.\
An important consequence of this difference is that the cut off distance $R_c^{}$
determined by ab initio MD ($1.30\pm 0.03$ \AA) must be adapted
 to that by ab initio PIMD 
 ($1.50\pm 0.02$ \AA  (H$_2$O) $1.53\pm 0.04$ \AA
 (D$_2$O),  see Table II or SI,  respectively) 
 in order to reproduce the reference \pkw\ value of liquid water.
We note in passing that the result obtained from ab initio MD is consistent 
with values used in previous studies of $1.22-1.3$ \AA.
\cite{davies_estimating_2002,doltsinis_theoretical_2003,
chen_prediction_2012,schilling_determination_2019}
One way of interpreting the increase in cut off distance in the PIMD simulations is
that the proton stays bonded to the restrained oxygen longer due to NQEs.\ 


In Figures \ref{fig::FreeEnerg}(b) and \ref{fig::FreeEnerg}(c) we display
 the results obtained from semiempirical MD/PIMD
 and empirical MD/PIMD, respectively.
Comparing Figures \ref{fig::FreeEnerg}(b) and \ref{fig::FreeEnerg}(c) with Figure \ref{fig::FreeEnerg}(a),
 the free energy curves of
 semiempirical MD and PIMD look very different from
 that of ab initio MD and PIMD in the absolute values.
Accordingly, the cutoff distances $R_c^{}$ are quite
 different from one another.
On the other hand, the reduction of the free energy curves behaves
 similarly with respect to the NQEs.
Therefore, we speculate that the isotope effects
 predicted by empirical and semiempirical PIMD
 can be as reliable as those from ab initio PIMD.


The free energy curves of the classical simulation show an
 anharmonic behavior since the mean force upon proton
 dissociation is a nonlinear function of $r$.
In addition, the difference between the free energy curves of the
 classical and quantum simulations vary along $r$,
 which implies the influence of anharmonicity on the
 NQEs.\
In general, the magnitude of the NQEs\ tends
 to be large where the potential curvature $\omega$
 is larger than $1/\beta\hbar$, which can be
 understood from the formula of the quantum harmonic
 correction to the free energy,
 $A_{\rm qhc}^{}
 = - \beta^{-1} \log\left\{ \frac{\beta\hbar\omega/2}
 {\sinh(\beta\hbar\omega/2)} \right\}$.\cite{shiga_quantum_2012}

\begin{figure}
\includegraphics[width=\linewidth]{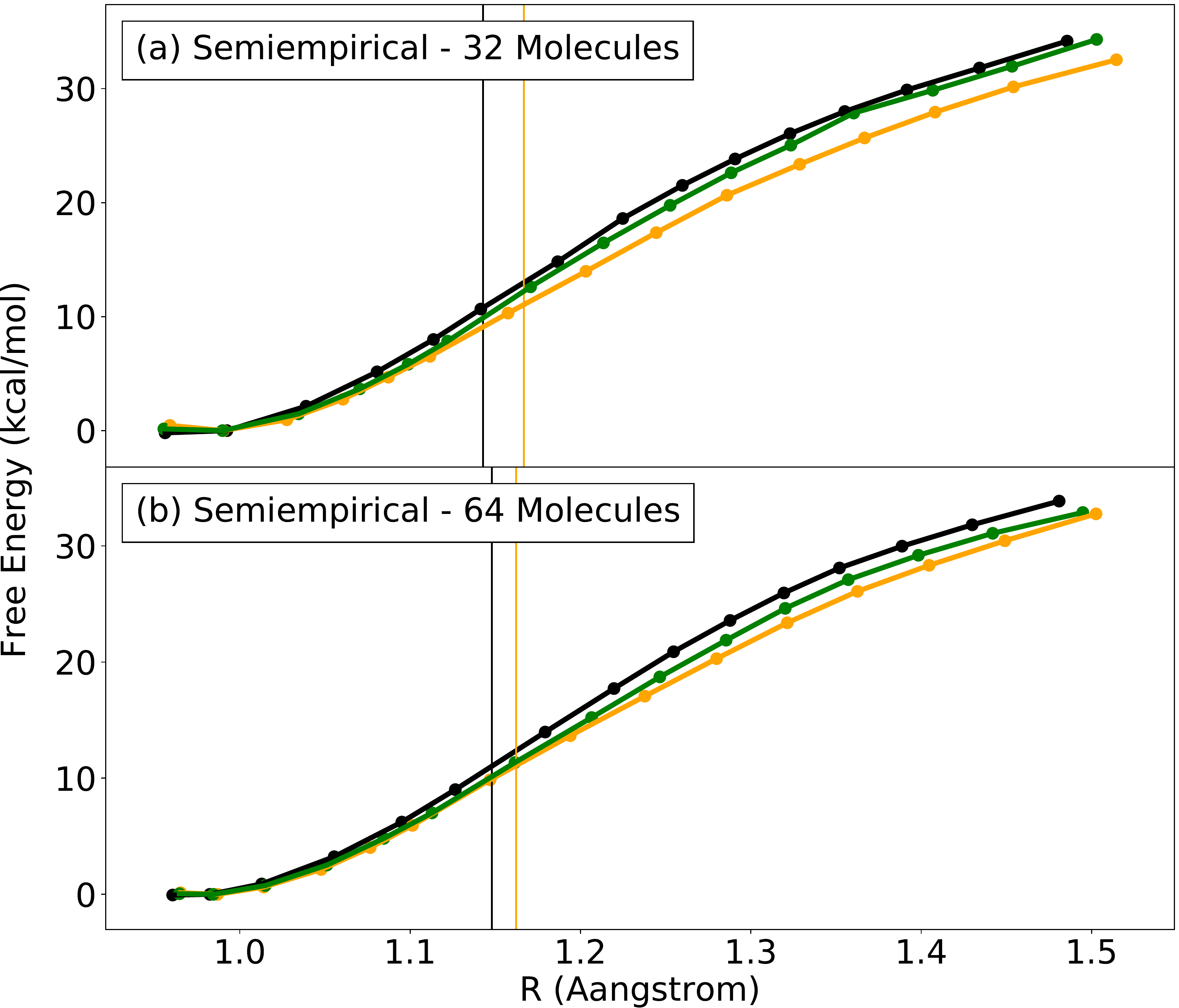}
\caption{}
\label{Figure::FreeEnergySizeComp}
\end{figure}

The computational effort needed to obtain the free energy curve increases
 proportionally with respect to the number of restraints.
We therefore setup the ab initio MD and PIMD simulations
 with a smaller system and number of beads ($N=32$ and $P=12$)
 compared to our previous work on the same system
 without restraints ($N=64$ and $P=16$).\cite{machida_nuclear_2017}
To verify this setup, we checked the size- and bead-dependence
 of the MD and PIMD simulations using the semiempirical
 DFTB potential.
Figure \ref{Figure::FreeEnergySizeComp}(a) shows the free energy curves 
obtained from semiempirical PIMD simulation for a larger number of beads 
with $P=32$, while Figure \ref{Figure::FreeEnergySizeComp}(b) shows the 
free energy curves obtained from semiempirical MD and PIMD simulation 
for a larger system size with 64  water molecules.
It can be clearly seen that Figures \ref{Figure::FreeEnergySizeComp}(a)
 and (b) follow the same trend as Figures \ref{fig::FreeEnerg}(b),
 in the sense that the free energy is reduced by the NQEs\
 within the semiempirical simulations.
We can therefore expect that the results obtained from
 ab initio MD and PIMD simulations shown in Figure 
 \ref{fig::FreeEnerg}(a) are reasonable with respect 
 to the nuclear quantum and isotope effects on 
 the free energy curves.


In Table II, we display the autoionization constants, \pkw,
 of liquid D$_2^{}$O and T$_2^{}$O calculated using Equation 
 (\ref{eq::pkwprop}) and the free energy curves obtained from 
 PIMD simulations.
The cutoff parameter $R_c^{}$ was determined
 for a particular setup of PIMD simulations
 such that the \pkw\ of liquid H$_2^{}$O is 14.00.
We see the trend that the \pkw\ value of liquid T$_2^{}$O is
 higher than 14.00, which is consistent with experimental expectations.
 \cite{ceriotti_nuclear_2016}\
However, the difference between the \pkw\ values of liquid
 H$_2^{}$O and D$_2^{}$O in the ab initio PIMD simulations
  lies within the error bars.
All the PIMD simulations were able to predict
 that the \pkw\ value of liquid T$_2^{}$O is larger than that of
 liquid H$_2^{}$O with statistical significance.
We also tried calculating \pkw\ using the D$_2$O \pkw\
as a reference, results given in Table SI of
the supplementary information (SI), and we found that
the results agree well with the ones presented in Table
II.


The autoionization constants calculated using the absolute 
method outlined around Equation (\ref{eq::pwkabs}) are given in
Table III.
This method makes it possible to compare the \pkw\ calculated
using MD and PIMD simulations directly, as no reference value
is required for this method.
Comparing the ab initio MD and PIMD results for H$_2^{}$O reveals
that the NQEs\ play an important role in determining the correct 
\pkw\ value.
We note that the values predicted from the semi empirical MD and PIMD
simulations are far from the correct value of \pkw, 
see Table\ SII in the SI, as can be expected from their free
energy profiles.
The results using the empirical OSS force field\ are, however, 
comparable with those of the ab initio method, with a slightly
smaller difference between MD and PIMD results.
As shown above, the absolute method of ab initio PIMD simulation 
can correct for the large overestimation of the unbinding energy of a proton 
from water of ab initio MD simulation. Additionally, the absolute method
produces a reasonable result for the isotope effect for all pure 
isotopologues studied here.


In Table IV, we display the acidity constants, \pka, associated
 with the reaction,
\begin{equation}
{\rm HXO}({\rm aq})
 \rightleftharpoons {\rm OX}^-({\rm aq}) + {\rm H}^+({\rm aq}),
\end{equation}
where ${\rm X = D}$ or ${\rm T}$.
The values were calculated using Equation (\ref{eq::pkapropext})
 and the free energy curves obtained from PIMD simulations.

Experimental values do not exist for the \pka\ of HDO and HTO.\
However, the ``rational'' \pka\ of H$_2$O, which is $15.74 = $
 \pkw$+\log\left(c_W \right)$,
 would be a reference.
The rational \pka\ is obtained
 by using an activity for the H$_2$O molecules
 when the dissociating water molecule is assumed misleadingly to be
 distinguishable from the rest of the water molecules in solution.
\cite{silverstein_pka_2017,meister_confusing_2014}
\
It can however serve well as a reference in this case where the HDO and HTO
molecules are in fact distinguishable from the solvent molecules.

It is expected that
 the \pka\ of HDO and HTO in water
 are either similar to or larger than the ``rational'' \pka\ of H$_2$O,
 assuming that the isotope effect
 follows the same trend as the case of the \pkw\ of
 H$_2$O, D$_2$O and T$_2$O.

Here it turned out that the

 \pka\ value of HDO predicted from empirical PIMD is close to
 the ``rational'' \pka\ of H$_2^{}$O, while the predicted \pka\ value of HTO
 is larger than that of H$_2^{}$O with statistical significance.
For the semiempirical calculations we find that both the \pka\
of HDO and HTO are larger than the ``rational'' \pka\ of H$_2^{}$O.
We note that these \pka\ values are difficult to measure
 experimentally, but they are important in determining
 the concentration of ${\rm OD}^-$ and ${\rm OT}^-$ ions
 in liquid water.

\section{Conclusion}
In this study we established a first-principles approach to compute
 the autoionization and acidity constants of water taking account of NQEs\ by PIMD with CN restraints. 
The simulations were carried out using different potential energy surfaces,
 \textit{i.e.}, ab initio DFT, semiempirical DFTB and empirical OSS2 methods.

The findings presented here are in line with previous empirical 
results.\cite{ceriotti_nuclear_2016}
The current study does differentiate itself from previous studies,  
by targeting the autoionization process directly,  without any 
empirical factors to estimate its free energy curves.


It was found that the free energy curve in the proton dissociation obtained from the 
quantum PIMD simulation is downshifted significantly compared with that obtained 
from the classical MD simulation, thus showing the importance of the NQEs\ on the autoionization 
and acidity constants. 
The \pkw\ values of water isotopologues, liquid D$_2^{}$O and T$_2^{}$O, were estimated based on a
probabilistic method using shifts in the free energy curves of D$_2^{}$O and T$_2^{}$O 
with respect to that of H$_2^{}$O. 
The results agree well with experimental values, accounting for the statistical uncertainties of our simulations.


We went on to compute the \pka\ values of aqueous HDO and HTO molecules which
are difficult to measure experimentally. The results predict that \pka\ of 
aqueous HTO (HDO) is larger than (close to) that of H$_2^{}$O.


The work presented here opens the possibility for accurate calculations 
of \pka\ for more complex systems, such as small organic molecules in solution.
It furthermore makes it possible to to predict the isotope effect on these system by
direct calculation.
These goals and calculating the temperature and pressure dependence of the autoionization constant of water
will be a\ subject of future studies.


We finally note that the probabilistic method requires the reference \pkw\ value of H$_2^{}$O (14.00)
 while the absolute method does\ not.
For the latter, however, an accurate estimation of absolute \pka\ values remains a difficult challenge.
In fact the \pkw\ estimated using the absolute method with the present 
simulations have very different values to those from the probablistic method, see Tables II and III. 
This is because the standard free energy curve is strongly dependent on the potential 
models and the system sizes. These issues should be studied more carefully in the future.
It is however clear that the inclusion of the NQEs\ is important for determining autoionization and acidity constants.

In fact, we do find a difference of $4.5\pm 0.9$ \pka\ units between the 
ab initio MD and PIMD results, where the PIMD result is clearly closer 
to the true value of the \pkw\ of water. 
This difference does seem to be in line, to some extent, with the experimental 
extrapolation of 3 \pka\ units which was suggested earlier.
\cite{ceriotti_nuclear_2016}


\section{Supplementary material}
 See the supplementary material for Tables showing
 the autoionization constants of water isotopologues calculated by
 the probabilistic method using the experimental
 \pkw\ of D$_2$O to calculate $R_c$,
 and the autoionization constants of water isotopologues
 calculated by the absolute method using the semiempirical DFTB method.

\section{Acknowledgements}
This work was completed under the project ``Hydrogenomics'' in Grant-in-Aid for Scientific Research on Innovative Areas, MEXT, Japan. The computations were mostly run on the supercomputer facilities at Japan Atomic Energy Agency and the Institute for Solid State Physics, The University of Tokyo. M.S. thanks JSPS KAKENHI (18H05519, 18H01693, 18K05208) and MEXT Program for Promoting Research on the Supercomputer Fugaku (Fugaku Battery \& Fuel Cell Project) for financial support. We thank Prof. Nikos Doltsinis in Universit\"at M\"unster for his advice on the coordination number constraints. We thank Dr. Alex Malins in JAEA for proofreading the text.

\section{Data Availability}

The data that support the findings of this study are available within the article and its supplementary material.

\bibliographystyle{ieeetr}
\bibliography{Paper.v5.bib}


\clearpage

\begin{center}
{\bf Figure Captions}
\end{center}
\ \\ \ \\
Figure 1: (A) Initial structure containing 32 water molcules. The four inserts on the right show the water molecules within 4 \AA\ of the central oxygen molecule during the simulation with a coordination number restraint of (B) 0.98, (C) 0.80, (D) 0.60, (E) 0.31. The distance from the central oxygen to the nearest constrained hydrogen is (B) 1.00 \AA, (C) 1.16 \AA, (D) 1.29 \AA, (E) 1.47 \AA. The O-H bonds are for all figures drawn if the O-H distance is less than 1.3 \AA, with the exception of the O$^*$-H bond in which case the bonds are drawn for distances up to 1.5 \AA. The central oxygen is in (B-E) marked with orange, and the teal hydrogen atom is the only hydrogen atom not subject to any constraints during the simulation.
\ \\ \ \\
Figure 2: The free energy curves, $A(n(r))$, for H$_2^{}$O using MD (black), H$_2^{}$O using PIMD (orange) and T$_2^{}$O using PIMD (green). (a) was calcultated using the ab initio potential, while (b) uses the semiempirical potential and (c) uses the empirical potential. For the MD simulations using OSS2, DFTB and DFT we found the values of $R_c$ to be $1.39\pm0.02$ \AA, $1.13\pm 0.00$ \AA, $1.27\pm 0.01$ \AA, respectively.  See Table\ II for the corresponding values for the PIMD simulations. The $R_c$ values are shown in this figure as vertical lines for H$_2^{}$O using MD (black) and PIMD (orange).
\ \\ \ \\
Figure 3: The free energy curves of semiempirical simulations for H$_2$O and T$_2$O in a box containing (a) 32 and (b) 64 molecules. In both figures the black curve represents classical MD on H$_2$O, and the corresponding black vertical line is the calculated $R_c$ distance for this simulation, (a) $1.13\pm 0.00$ \AA\ and (b) $1.15\pm 0.00$ \AA, respectively. The orange and green curves represent H$_2$O and T$_2$O respectively in a PIMD simulation with (a) P = 32 or (b) P = 12. The orange vertical lines represent the calculated $R_c$ of the two H$_2$O PIMD simulations, (a) $1.16\pm 0.00$ \AA\ and (b) $1.17\pm 0.00$ \AA, respectively.


\clearpage

\clearpage

\clearpage

\clearpage

\begin{table*}[htp]
  \begin{center}
     \begin{tabular}{ccccccc}
        \multicolumn{7}{l}
        {Table I. Simulation setup of
         water isotopologues.} \\
        \hline\hline
            \hspace{0mm} Method          \hspace{0mm}
          & \hspace{0mm} System          \hspace{0mm}
          & \hspace{0mm} Molecules       \hspace{0mm}
          & \hspace{0mm} $P$             \hspace{0mm}
          & \hspace{0mm} $\Delta t$ [ps] \hspace{0mm}
          & \hspace{0mm} Length [ps]     \hspace{0mm}
          & \hspace{0mm} Restraints      \hspace{0mm} \\
        \hline
            DFT$^a$  & liq. H$_2^{}$O & 32 &  1 & 0.25 & 24.0 & 15 \\
            DFT$^b$  & liq. H$_2^{}$O & 32 & 12 & 0.25 & 25.0 & 15 \\
            DFT$^b$  & liq. D$_2^{}$O & 32 & 12 & 0.25 & 25.0 & 15 \\
            DFT$^b$  & liq. T$_2^{}$O & 32 & 12 & 0.25 & 25.0 & 15 \vspace*{3mm} \\

            DFTB$^c$ & liq. H$_2^{}$O & 32 &  1 & 0.25 & 25.0 & 15 \\
            DFTB$^d$ & liq. H$_2^{}$O & 32 & 12 & 0.25 & 25.0 & 15 \\
            DFTB$^d$ & liq. H$_2^{}$O & 32 & 32 & 0.25 & 25.0 & 15 \\
            DFTB$^d$ & liq. D$_2^{}$O & 32 & 12 & 0.25 & 25.0 & 15 \\
            DFTB$^d$ & liq. T$_2^{}$O & 32 & 12 & 0.25 & 25.0 & 15 \\
            DFTB$^d$ & liq. T$_2^{}$O & 32 & 32 & 0.25 & 25.0 & 15 \\            
            DFTB$^d$ & aq.  HDO       & 32 & 12 & 0.25 & 25.0 & 15 \\
            DFTB$^d$ & aq.  HTO       & 32 & 12 & 0.25 & 25.0 & 15 \vspace*{3mm} \\
            DFTB$^c$ & liq. H$_2^{}$O & 64 &  1 & 0.25 & 25.0 & 15 \\
            DFTB$^d$ & liq. H$_2^{}$O & 64 & 12 & 0.25 & 25.0 & 15 \\
            DFTB$^d$ & liq. T$_2^{}$O & 64 & 12 & 0.25 & 25.0 & 15 \vspace*{3mm} \\
            OSS2$^e$ & liq. H$_2^{}$O & 64 &  1 & 0.25 & 12.5 & 15 \\
            OSS2$^f$ & liq. H$_2^{}$O & 64 & 12 & 0.25 & 12.5 & 15 \\
            OSS2$^f$ & liq. D$_2^{}$O & 64 & 12 & 0.35 & 17.5 & 15 \\
            OSS2$^f$ & liq. T$_2^{}$O & 64 & 12 & 0.43 & 21.5 & 15 \\
            OSS2$^f$ & aq.  HDO       & 64 & 12 & 0.25 & 12.5 & 15 \\
            OSS2$^f$ & aq.  HTO       & 64 & 12 & 0.25 & 12.5 & 15 \\
        \hline
       \multicolumn{7}{l}{$^a$Ab initio MD.} \\
       \multicolumn{7}{l}{$^b$Ab initio PIMD.} \\
       \multicolumn{7}{l}{$^c$Semiempirical MD.} \\
       \multicolumn{7}{l}{$^d$Semiempirical PIMD.} \\
       \multicolumn{7}{l}{$^e$Empirical MD.} \\
       \multicolumn{7}{l}{$^f$Empirical PIMD.} \\
     \end{tabular}
   \end{center}
\end{table*}

\clearpage

\begin{table}[htp]
  \begin{center}
     \begin{tabular}{cccccc}
        \multicolumn{6}{l}
         {Table II. Autoionization constants of water
          isotopologues\ calculated by} \\
           \multicolumn{6}{l}
         {the probabilistic method.} \\
        \hline\hline
            \hspace{3mm} Method \hspace{3mm}
          & \hspace{3mm} System \hspace{3mm}
          & \hspace{0mm} Molecules       \hspace{0mm}
          & \hspace{0mm} $P$             \hspace{0mm}
          & \hspace{3mm} \pkw\ (PROB) \hspace{3mm}
          & \hspace{3mm} $R_c^{}$ [\AA] \hspace{3mm} \\
        \hline
           DFT$^a$  & liq. H$_2$O & 32 & 12 & 14.0$^e$       & 1.50$\pm$0.02 \\
           DFT$^a$  & liq. D$_2$O & 32 & 12 & 13.5 $\pm$ 1.2 & 1.50$\pm$0.02 \\
           DFT$^a$  & liq. T$_2$O & 32 & 12 & 15.4 $\pm$ 0.9 & 1.50$\pm$0.02 \\
        \\
           DFTB$^b$ & liq. H$_2$O & 32 & 12 & 14.0$^e$       & 1.16$\pm$0.00 \\
           DFTB$^b$ & liq. D$_2$O & 32 & 12 & 15.0 $\pm$ 0.3 & 1.16$\pm$0.00 \\
           DFTB$^b$ & liq. T$_2$O & 32 & 12 & 15.0 $\pm$ 0.2 & 1.16$\pm$0.00 \\
        \\
           DFTB$^b$ & liq. H$_2$O & 32 & 32 & 14.0$^e$       & 1.17$\pm$0.00 \\
           DFTB$^b$ & liq. T$_2$O & 32 & 32 & 15.9 $\pm$ 0.3 & 1.17$\pm$0.00 \\
        \\
           DFTB$^b$ & liq. H$_2$O & 64 & 12 & 14.0$^e$       & 1.16$\pm$0.00 \\
           DFTB$^b$ & liq. T$_2$O & 64 & 12 & 14.4 $\pm$ 0.2 & 1.16$\pm$0.00 \\
        \\
           OSS2$^c$ & liq. H$_2$O & 64 & 12 & 14.0$^e$       & 1.51$\pm$0.01 \\
           OSS2$^c$ & liq. D$_2$O & 64 & 12 & 15.3 $\pm$ 0.5 & 1.51$\pm$0.01 \\
           OSS2$^c$ & liq. T$_2$O & 64 & 12 & 15.6 $\pm$ 0.5 & 1.51$\pm$0.01 \\
        \\
           Exptl$^d$ & liq. H$_2$O & 14.00         &                \\
           Exptl$^d$ & liq. D$_2$O & 14.86\cite{shoesmith_ionization_1976}  &                \\
           Exptl$^d$ & liq. T$_2$O & 15.2\cite{ceriotti_nuclear_2016}       &                \\
       \hline
         \multicolumn{6}{l}
           {$^a$Ab initio PIMD.} \\
         \multicolumn{6}{l}
           {$^b$Semiempirical PIMD.} \\
         \multicolumn{6}{l}
           {$^c$Empirical PIMD.} \\
         \multicolumn{6}{l}
           {$^d$Experimental values.} \\
         \multicolumn{6}{l}
           {$^e$Reference value to determine $R_c^{}$.}
     \end{tabular}
   \end{center}
\end{table}


\clearpage
\normalsize

\begin{table}[htp]
  \begin{center}
     \begin{tabular}{ccccc}
        \multicolumn{5}{l}
         {Table III. Autoionization constants of water
          isotopologues\ calculated by} \\
        \multicolumn{5}{l}
         {the absolute method.} \\
        \hline\hline
            \hspace{3mm} Method \hspace{3mm}
          & \hspace{3mm} System \hspace{3mm}
          & \hspace{0mm} Molecules       \hspace{0mm}
          & \hspace{0mm} $P$             \hspace{0mm}
          & \hspace{3mm} \pkw\ (ABS) \hspace{3mm}\\
        \hline
           DFT$^a$  & liq. H$_2$O & 32 & 1 & 19.7 $\pm$ 0.3 \\
           DFT$^b$  & liq. H$_2$O & 32 & 12 & 15.2 $\pm$ 0.9 \\
           DFT$^b$  & liq. D$_2$O & 32 & 12 & 14.8 $\pm$ 0.7  \\
           DFT$^b$  & liq. T$_2$O & 32 & 12 & 16.0 $\pm$ 0.8  \\
        \\
           OSS2$^c$ & liq. H$_2$O & 64 & 1 & 15.0 $\pm$ 0.5 \\
           OSS2$^d$ & liq. H$_2$O & 64 & 12 & 13.5 $\pm$ 0.5 \\
           OSS2$^d$ & liq. D$_2$O & 64 & 12 & 14.9 $\pm$ 0.7 \\
           OSS2$^d$ & liq. T$_2$O & 64 & 12 & 14.6 $\pm$ 0.6 \\
       \hline
         \multicolumn{5}{l}
           {$^a$Ab initio MD.} \\
         \multicolumn{5}{l}
           {$^b$Ab initio PIMD.} \\
         \multicolumn{5}{l}
           {$^c$Empirical MD.} \\
         \multicolumn{5}{l}
           {$^d$Empirical PIMD.}
     \end{tabular}
   \end{center}
\end{table}


\clearpage

\begin{table}[htp]
  \begin{center}
     \begin{tabular}{cccc}
        \multicolumn{4}{l}
        {Table IV. Acidity constants of water
         isotopologues,  calculated using Equation (23).} \\
        \hline\hline
            \hspace{3mm} Method \hspace{3mm}
          & \hspace{3mm} System \hspace{3mm}
          & \hspace{3mm} \pka\ (PROP)\hspace{3mm}
          & \hspace{3mm} $R_c^{}$ [\AA] \hspace{3mm} \\
        \hline
            DFTB$^a$  & aq. HDO & 16.1 $\pm$ 0.1 & 1.16$\pm$0.00 \\
            DFTB$^a$  & aq. HTO & 16.3 $\pm$ 0.1 & 1.16$\pm$0.00 \\
            OSS2$^b$  & aq. HDO & 15.6 $\pm$ 0.3 & 1.51$\pm$0.01 \\
            OSS2$^b$  & aq. HTO & 17.1 $\pm$ 0.4 & 1.51$\pm$0.01 \\
        \hline
         \multicolumn{4}{l}
           {$^a$Semiempirical PIMD of 32 water molecules with $P=12$.} \\
         \multicolumn{4}{l}
           {$^b$Empirical PIMD of 64 water molecules with $P=12$.} 
     \end{tabular}
   \end{center}
\end{table}

\end{document}